\documentclass{article}
\usepackage{amssymb}

\usepackage{amsmath}
\usepackage{graphicx}
\usepackage{float}
\usepackage{color}

\input{tcilatex} 
\begin{document}

\title{Pinning of a drop by a junction on an incline}
\author{Jo\"{e}l De Coninck$^{1}$, Juan Carlos Fern\'{a}ndez Toledano$^{1}$ \and Fran%
\c{c}ois Dunlop$^{2}$, Thierry Huillet$^{2}$ \\
$^{1}$Laboratoire de Physique des Surfaces et Interfaces,\\
Universit\'{e} de Mons, 20 Place du Parc, \\
7000 Mons, Belgium\\
$^{2}$Laboratoire de Physique Th\'{e}orique et Mod\'{e}lisation,\\
CNRS-UMR 8089, Universit\'{e} de Cergy-Pontoise,\\
95302 Cergy-Pontoise, France}
\maketitle

\begin{abstract}
The shape of a drop pinned on an inclined {substrate} is a long-standing
problem where the complexity of real surfaces, with heterogeneities and
hysteresis, makes it complicated to understand the mechanisms behind the
phenomena. Here we consider the simple case of a drop pinned on an incline
at the junction between a hydrophilic half-plane (the top half) and a
hydrophobic one (the bottom half). Relying on the equilibrium equations
deriving from the balance of forces, we exhibit three scenarii depending on
the way the contact line of the drop on the substrate either simply leans
against the junction or overfills (partly or fully) into the hydrophobic
side. We draw some conclusions on the geometry of the overlap and the
stability of these tentative equilibrium states. In the corresponding
retention force factor, we find that a major role is played by the wetted
length of the junction line, in the spirit of Furmidge's observations. The
predictions of the theory are compared with extensive molecular dynamics
simulations.
\end{abstract}

\section{Introduction}

As first described by Thomas Young \cite{Young1805} in his essay on cohesion
of fluids in 1805, the competition between the cohesion of a fluid to itself
and its adhesion to a solid gives rise to an angle of contact $\theta _{0}$
between the liquid and the solid that is specific to a given system at
equilibrium. This is now well known as the Young equation:

\begin{equation}
\gamma \cos \theta _{0}=\gamma _{SV}-\gamma _{SL}  \label{eq:young}
\end{equation}
where $\theta _{0}$ is the equilibrium contact angle and $\gamma _{SV}$ and $%
\gamma _{SL}$ are the Solid-Vapor and Solid-Liquid surface tensions,
respectively. It has been proven recently that this equation holds down to
the nanometric scale \cite{Das2011, Seveno2013b, Toledano2017}. {In practice
however, this equation holds for pure liquids on flat glasses or silica
wafers}. For real heterogeneous surfaces, chemically or physically, the
situation is more complex. We have to introduce the advancing ($\theta ^{A}$%
), the receding ($\theta ^{R}$) static contact angles and the difference
between both, which is called the hysteresis and arises from surface
roughness and/or heterogeneity \cite{Yarnold1946, Gregg1948, Johnson1964,
Joanny1984, Schwartz1985}. The contact angle of a sessile drop actually
observed will lie between $\theta ^{A}$ to $\theta ^{R}$ and is function of
the process of reaching the particular equilibrium state. See \cite{dGBQ}
for background and references.

The variety of possible processes and motions makes the prediction of the
final static contact angle challenging. No generally applicable correlation
between hysteresis and roughness features is known for a given surface. When
the corresponding substrate is tilted by a small angle $\alpha $, the drop
usually deforms its shape but remains pinned on the substrate. It is only
when the tilt angle $\alpha $ becomes large enough, above the value $\alpha
_{c}$, that the drop starts to slide. It has been proposed by Furmidge \cite
{Furm}, Eq.~$5,$ that

\begin{equation}
mg\sin \alpha _{c}=\frac{k}{2}w\gamma (\cos \theta ^{R}-\cos \theta ^{A})
\label{eq:Fermigier}
\end{equation}
where $m$ is the mass of the drop, $g$ the gravity constant, $\theta ^{R}$
and $\theta ^{A}$ the receding and the advancing contact angles, $w$ the
width of the drop in the direction perpendicular to inclination. The
dimensionless retention-force factor $k$ is close to $2$ according to \cite
{Furm}, Table $2$, but its value has been reexamined since then (\cite{BOS}
Eq.~$27$; \cite{El1} Eq.~$1$, \cite{El2} Eq.~$1$, \cite{Santos} Eq.~$3$, 
\cite{WSRV} Eq.~$4$ and \cite{CDH17} Eqs.~$1$ and $2$), concluding to $k$
varying in the range $\pi /2\leq k\leq 2$, depending on the physical
situation. Several studies have been devoted to this equation through
experiments \cite{BT}, numerical or theoretical calculations \cite{Sem}.
Mostly, all these studies differ by their hypothesis concerning the shape of
the contact line or different conditions for the experiments.

We are herewith studying the basic case where there is a chemical step in
the substrate. Experimentally this is a difficult situation simply because
the difference of wettability will be associated to a zone and not to a
line. To explore in details the validity of equations like Eq. (\ref
{eq:Fermigier}) avoiding unnecessary hypothesis, it is interesting to
revisit this problem using large scale molecular dynamics.\newline

The problem of a drop on an incline at the junction between a hydrophilic
half-plane and a hydrophobic half-plane has been addressed previously,
notably by simulation with Surface Evolver. See \cite{Be}, \cite{BB} and
references therein. An inclined chemical step has also been considered in 
\cite{Semp}. The incline has also been replaced by a wettability gradient 
\cite{Mou}.

{The paper is organized as follows. The theoretical aspects are given in
Section 2. Then we present the corresponding molecular dynamics simulations
in Section 3. Section 4 is devoted to a comparison between the two
approaches. Finally, some concluding remarks are presented in Section 5.}

\section{Drop on incline at hydrophilic-hydrophobic junction: theory}

\subsection{\label{junctionT}Drop pinned on inhomogeneous incline}

\begin{figure}[tbp]
\begin{center}
\resizebox{8cm}{!}{\includegraphics{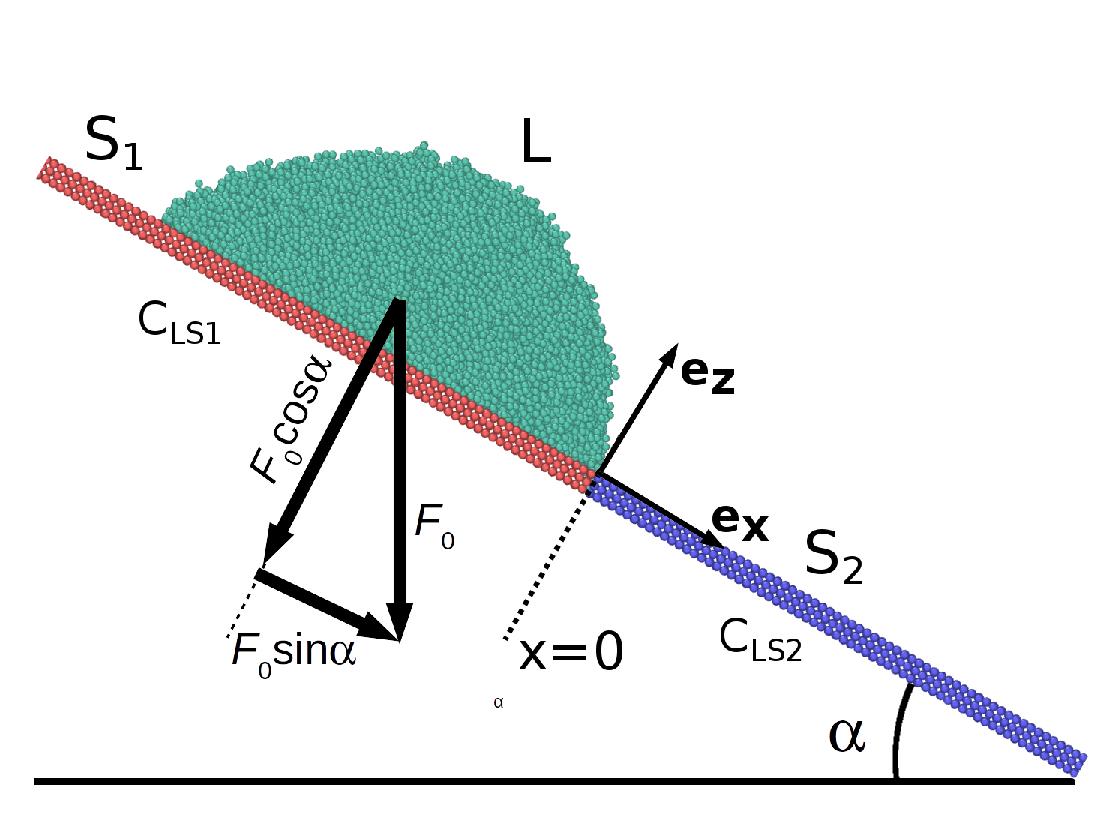}}
\end{center}
\caption{Scheme of the liquid drop on incline plane close to the boundary
between two solids under an external force $F_{0}$. }
\label{fig:scheme}
\end{figure}

{We first study the case of a drop pinned on an arbitrary inhomogeneous
incline. While considering the total capillary force, the gravity force and
the total pressure force acting upon the drop sum up to zero at equilibrium,
we obtain, while projecting on each axis relative to the incline, a system
of three equations relating the gravity force components to contour
integrals along the contact line and with a pressure contribution in the
direction perpendicular to the incline. In more details} we consider an
incline of angle $\alpha $ with respect to the horizontal. The $x$-axis is
along the slope downhill, the $y$-axis is horizontal, the $z$-axis is
perpendicular to the incline. The corresponding orthonormal basis is $%
\mathbf{e_{x}},\mathbf{e_{y}},\mathbf{e_{z}}$ as in Fig. \ref{fig:scheme}.
The basis of the drop is denoted $\Omega $ and its boundary, the contact
line $\partial \Omega $, is assumed piecewise differentiable, and $\mathbf{n}
$ is the outer normal to the contact line in the substrate plane. The
contact angle $\theta $ is assumed piecewise continuous on $\partial \Omega $%
. The total capillary force $\mathbf{F}$ upon the drop is 
\begin{equation}
\mathbf{F}=\gamma \,\oint_{\partial \Omega }\,dl\,(\mathbf{n}\cos \theta -%
\mathbf{e_{z}}\sin \theta )  \label{capf}
\end{equation}
The gravity force upon the drop is 
\[
m\mathbf{g}=(mg\sin \alpha ,\,0,\,-mg\cos \alpha ) 
\]
The total pressure force upon the drop is 
\begin{equation}
\mathbf{e_{z}}\int_{\Omega }dxdy\left( p(x,y,0)-p_{\mathrm{atm}}\right)
\end{equation}
These formulae remain valid if the contact line has a slow motion so that
the drop profile is always in the equilibrium shape conditioned by the
instantaneous contact line. The pinning and depinning of the contact line
depend upon $\gamma _{SV}-\gamma _{SL}$, where $\gamma _{SV}$ and $\gamma
_{SL}$ are the local Solid-Vapor (air) and Solid-Liquid (water) surface
tensions, which typically are not smooth functions, and which play a role in
(\ref{capf}) only through the choice of the contact line $\partial \Omega $.
Therefore, in these formulae, the contact angle $\theta $ can be any angle
between the local advancing ($\theta ^{A}$) and receding ($\theta ^{R}$)
angles.

At equilibrium the forces upon the drop sum up to zero, on each axis: 
\begin{eqnarray}
0 &=&mg\sin \alpha +\gamma \,\oint_{\partial \Omega }\,dl\,\mathbf{n}\cdot 
\mathbf{e_{x}}\,\cos \theta  \label{eqx} \\
0 &=&\oint_{\partial \Omega }\,dl\,\mathbf{n}\cdot \mathbf{e_{y}}\,\cos
\theta  \label{eqy} \\
0 &=&-mg\cos \alpha -\gamma \,\oint_{\partial \Omega }dl\,\sin \theta
+\int_{\Omega }dxdy\left( p(x)-p_{\mathrm{atm}}\right)  \label{eqz}
\end{eqnarray}
where $p(x)=p(x,y,0)=p(0,0,0)+\rho gx\sin \alpha $ from the law of
hydrostatics. Equation (\ref{eqy}) will be automatically satisfied for a
drop symmetric with respect to the plane ${y=0}$, where $\mathbf{n}%
(x,\,-y)\cdot \mathbf{e_{y}}=-\mathbf{n}(x,\,y)\cdot \mathbf{e_{y}}$ and $%
\theta (x,-y)=\theta (x,y)$. 
\begin{figure}[tbp]
\begin{center}
\resizebox{12cm}{!}{\includegraphics{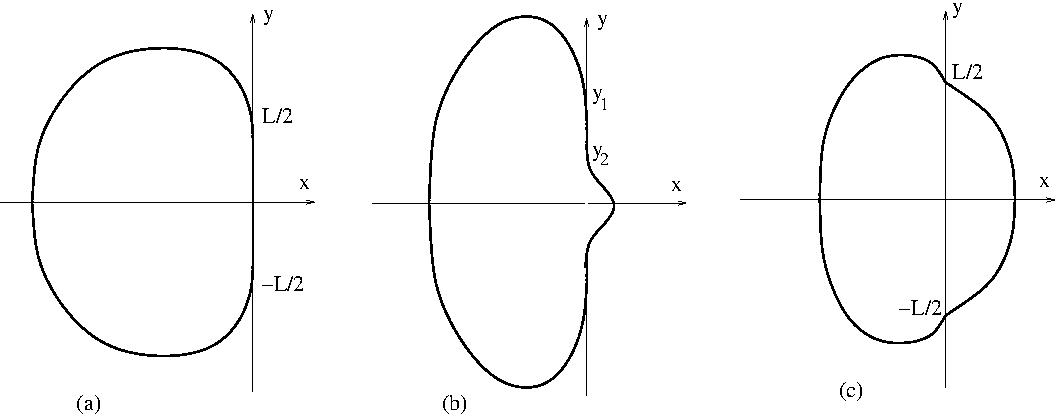}}
\end{center}
\caption{Contact line at hydrophilic-hydrophobic junction, gravity increased
from (a) to (c): (a) stable equilibrium, Eq. (\ref{eqa}); (b) tentative
equilibrium, Eq. (\ref{eqb}); (c) unstable equilibrium, Eq. (\ref{eqc}).}
\label{figL2}
\end{figure}

\subsection{Drop pinned at hydrophilic-hydrophobic junction}

{We next consider the same problem but specifically for a drop pinned on an
incline at the junction between a hydrophilic half-plane (the top half) and
a hydrophobic one (the bottom half). In this context, we discuss three
scenarii: $\left( a\right) $ one for which the contact line partly follows
the junction line on a segment of width $L$, $\left( b\right) $ one for
which part of the contact line goes into the hydrophobic half-plane in a
central protuberance while keeping two side overlaps with the junction line
and $\left( c\right) $ one for which part of the contact line crosses
straight into the hydrophobic half-plane. See Fig.~\ref{figL2}. }

{The three cases may be viewed as the ones obtained while successively
increasing the angle of the incline, or gravity or the volume of the drop,
in short increasing the Bond number $Bo$. The first case can be a stable
equilibrium, the second might be a stable equilibrium and the third is an
unstable equilibrium. }

{In this hydrophilic/hydrophobic junction setup, we derive the new version
of the equilibrium equations deriving from the balance of forces. For case $%
\left( a\right) $, equation (\ref{eqx}) along the axis of the slope downhill
equates the projection of the gravity force to a simple integral of the
cosine of the contact angle over the overlap segment of unknown length $L$;
observing that the contact angle is maximal in the middle of the overlap
segment and bounded above by the Young angle in the hydrophobic side, a
lower bound of $L$ is supplied. For case $\left( b\right) $, the same
equation equates the projection of the gravity force to the sum of two
contributions: one similar to the previous one but restricted to the overlap
junction line/contact line and one relative to the protuberance involving
the difference between the Young angle cosines in the
hydrophilic/hydrophobic half-planes times the length of the junction line
covered by the protuberance. Finally, for case $\left( c\right) $, there is
only one contribution to the balance equation involving the difference
between the Young angle cosines in the hydrophilic/hydrophobic half-planes
times the full length of the junction line covered by the drop. Let us now
formulate and justify this in details.}

The upper half-plane $\{x<0\}$ is a relatively hydrophilic substrate, of
Young angle $\theta _{1}$, while the lower half-plane $\{x>0\}$ is less
hydrophilic, of Young angle $\theta _{2}>\theta _{1}$. The contact line is
either $\partial \Omega =\partial \Omega _{1}\cup \partial \Omega _{12}$
where $\partial \Omega _{1}\subset \{x<0\}$ and $\partial \Omega _{12}=[-{%
\frac{L}{2}},{\frac{L}{2}}]\subset \{x=0\}$ for some $L>0$, see Fig.~\ref
{figL2}(a), or else it may be $\partial \Omega =\partial \Omega _{1}\cup
\partial \Omega _{2}\cup \partial \Omega _{12}$ where $\partial \Omega
_{2}\subset \{x>0\}$ and $\partial \Omega _{12}=[-y_{1},-y_{2}]\cup
[y_{2},y_{1}]\subset \{x=0\}$ for some $0<y_{2}\le y_{1}$, see Fig.~\ref
{figL2}(b). On $\partial \Omega _{1}$ the contact angle is the Young angle $%
\theta _{1}$ (ideal substrate, no hysteresis), on $\partial \Omega _{2}$ the
contact angle is the Young angle $\theta _{2}$ (ideal substrate, no
hysteresis). The contact angle along $\partial \Omega _{12}$ is a continuous
function $\theta (y)$ with $\theta _{1}\le \theta (y)\le \theta _{2}$. The
drop is symmetric under $y\to -y$, and the function $\theta (y)$ is
decreasing on $\partial \Omega _{12}\cap \{y\ge 0\}$.

Case (a) in Fig.~\ref{figL2}, $\partial \Omega _{2}=\emptyset $ certainly
occurs, by continuity, for small $\alpha $, with $L=L(\alpha )$ increasing
with $\alpha $ from $L(0)=0$. For small $\alpha $, the function $\theta (y)$
is independent of $\theta _{2}$ and reaches a maximum $\theta (0)<\theta
_{2} $.

Denoting $\mathbf{t}$ the unit tangent vector and $\mathcal{R}_{\pi /2}$ a
rotation by $\pi /2$, we have 
\begin{equation}
\oint_{\partial \Omega }\,dl\,\mathbf{n}=-\mathcal{R}_{\pi
/2}\oint_{\partial \Omega }\,dl\,\mathbf{t}=-\mathcal{R}_{\pi
/2}\oint_{\partial \Omega }\,d\mathbf{r}=0
\end{equation}
which allows to write (\ref{eqx}) as 
\[
0=mg\sin \alpha +\gamma \,\oint_{\partial \Omega }\,dl\,\mathbf{n}\cdot 
\mathbf{e_{x}}\,(\cos \theta -\cos \theta _{1}) 
\]
\begin{equation}
=mg\sin \alpha +\gamma \,\int_{-L/2}^{L/2}dy\,(\cos \theta -\cos \theta _{1})
\label{eqa}
\end{equation}
which implies 
\begin{equation}
L>\frac{mg\sin \alpha }{\gamma \,(\cos \theta _{1}-\cos \theta _{2})}
\end{equation}
Upon increasing $\alpha $ or $g$ or $m$, the configuration with $\partial
\Omega _{2}=\emptyset $ becomes unstable when $\theta (0)$ reaches $\theta
_{2}$, with a transition to $\partial \Omega _{2}\ne \emptyset $, Fig.~\ref
{figL2}(b). One may expect a continuous transition, with $y_{2}$ small at
the onset.

For the part of the contact line on the hydrophobic side, we have 
\begin{equation}
\int_{x>0}\,dl\,\mathbf{n}=-\mathcal{R}_{\pi /2}\int_{x>0}\,d\mathbf{r}=-%
\mathcal{R}_{\pi /2}2y_{2}\mathbf{e_{y}}=2y_{2}\mathbf{e_{x}}
\end{equation}
which allows to write (\ref{eqx}) as 
\[
0=mg\sin \alpha +\gamma \,\oint_{\partial \Omega }\,dl\,\mathbf{n}\cdot 
\mathbf{e_{x}}\,(\cos \theta -\cos \theta _{1}) 
\]
\begin{equation}
=mg\sin \alpha +2\gamma y_{2}(\cos \theta _{2}-\cos \theta _{1})+2\gamma
\int_{y_{2}}^{y_{1}}dy\,(\cos \theta -\cos \theta _{1})  \label{eqb}
\end{equation}

A second transition, perhaps the roll-off, may be expected when $y_{2}$
approaches $y_{1}$.

In the case of Fig.~ \ref{figL2}(c), Equation (\ref{eqx}) {remarkably
simplifies} and takes the form 
\begin{equation}
0=mg\sin \alpha +\gamma L\,(\cos \theta _{2}-\cos \theta _{1}).  \label{eqc}
\end{equation}
{This corresponds to the Furmidge formula (\ref{eq:Fermigier}) with }$\alpha 
$\ any angle less or equal $\alpha _{c},$ with $k=2,\emph{\ }$and $L$ {%
taking the role of the width }$w$ and the angles $\theta _{1},$ $\theta _{2}$
taking the roles of the receding and advancing angles $\theta ^{R}$, $\theta
^{A}.$

\subsection{Smoothness of equilibrium contact lines}

{{We herewith discuss the question of the smoothness of the contact line
where the contact line meets the junction line. By smoothness, it is meant
that the tangent vector to the contact line is continuous all along the
contact line. We give strong arguments in favor of smoothness showing that a
discontinuity would violate that the surface, as a solution of the
Laplace-Young equation, must have a bounded mean curvature. Counter-examples
of ``quasi-corners'' that cannot be droplet equilibrium shapes are supplied.
Corners and cusps of the contact line have been observed in moving droplets, 
\cite{Pod}, not in equilibrium.}}\newline

The contact lines shown on Fig.~\ref{figL2} may be smooth, with $\mathbf{t}$
and $\mathbf{n}$ continuous everywhere, or perhaps $\mathbf{t}$ and $\mathbf{%
n}$ could have jumps at $y=\pm L/2$ (Fig.~\ref{figL2},(a) or $y=\pm
y_{1},\pm y_{2}$ (Fig.~\ref{figL2},(b)). Also the contact line could cross
the junction with the contact angle jumping from $\theta _{1}$ to $\theta
_{2}$, as in Fig.~\ref{figL2},(c). An argument in favor of smoothness of the
contact line and continuity of the contact angle goes as follows. For
definiteness consider Fig.~\ref{figL2}(a), and assume $0<\theta _{1}<\pi $.
Consider the drop as a three-dimensional body, solution of the Laplace-Young
equation with the given boundary conditions. 
The drop surface is smooth except on the contact line: the tangent plane
below the drop is the substrate plane; the tangent plane on the liquid vapor
interface is well defined, and has a limit of slope $\tan \theta $ at the
contact line wherever the contact angle is well defined: everywhere except
possibly at $y=\pm L/2$. On the contact line, except possibly at $y=\pm L/2$%
, there are exactly two limiting tangent planes, limits from above with
slope $\tan \theta $ and from below with slope 0 (the slopes are defined
with respect to the $xy$-plane). The contact line is a sharp \textsl{edge}.

Now assume that $\mathbf{t}$ and $\mathbf{n}$ are discontinuous at $y=\pm
L/2 $. Then at this point there are three limiting tangent planes,
corresponding to the limits from below (slope 0), from above along the
contact line on the $x<0$ side (slope $\tan \theta _{1},\,\mathbf{n}\ne 
\mathbf{e_{x}}$) and from above along the contact line on the $x=0$ side
(slope $\tan \theta (L/2),\,\mathbf{n}=\mathbf{e_{x}}$). Let us call \textsl{%
quasi-corner} such a point with three limiting tangent planes but only two
limiting edges. 
\begin{figure}[tbp]
\begin{center}
\resizebox{10cm}{!}{\includegraphics{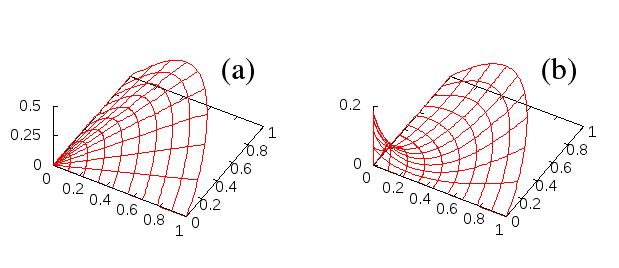}}
\end{center}
\caption{Quasi-corners: (a) $h(r,\varphi )=r\sin \varphi \cos \varphi $; (b) 
$h(r,\varphi )=(r\log r+1/e)\sin \varphi \cos \varphi $.}
\label{corner}
\end{figure}
For definiteness let us take the basis of the drop at the corner as $%
\{(x,y)\in [0,\infty )\times [0,\infty )\}$, as on Figs.~\ref{corner} and
use also polar coordinates with $x=r\cos \varphi ,\,y=r\sin \varphi $. Let $%
\mathbf{N}$ denote the normal vector to the fluid surface, which is well
defined except at the corner, here the origin. Let $\mathbf{N}_{1}$ be the
limit of $\mathbf{N}$ when approaching the origin along the $x$-axis and $%
\mathbf{N}_{2}$ the limit of $\mathbf{N}$ when approaching the origin along
the $y$-axis. For simplicity we take $\mathbf{N}_{1}$ and $\mathbf{N}_{2}$
constant along the corresponding axis, like equilibrium Young angles against
two different but homogeneous substrates. Consider a geodesic on the fluid
surface going from $(\epsilon ,0,0)$ to $(0,\epsilon ,0)$ and denote $s$ is
the corresponding curvilinear abscissa. On this path 
\begin{equation}
\mathbf{N}_{2}-\mathbf{N}_{1}=\int ds\,{\frac{d\mathbf{N}}{ds}}
\end{equation}
Assuming the drop surface continuous on the closed domain, including the
origin, the total length of the path is less than $2\epsilon $ and thus goes
to zero as $\epsilon $ goes to zero, while the left-hand-side is constant.
Therefore $d\mathbf{N}/ds$ must go to infinity. If $r\sim \epsilon $ along
the geodesic, 
then $d\mathbf{N}/ds\sim 1/r$. At least one of the principal curvatures is $%
\mathcal{O}(1/r)$. 
But the surface is a solution of the Laplace-Young equation, so that the
mean curvature is bounded. Therefore the other principal curvature must be
also $\mathcal{O}(1/r)$, with opposite sign in order to cancel the
divergence as $r\searrow 0$. The radial direction is likely to coincide with
the corresponding principal direction, orthogonal or near to the orthogonal
to the geodesic. Therefore $d\mathbf{N}/dr\sim 1/r$, or, for each $\varphi $%
, 
\begin{equation}
(1+h^{\prime }(r)^{2})^{-3/2}h^{\prime \prime }(r)\sim 1/r  \label{eqq}
\end{equation}
Taking the primitives on both sides: 
\begin{equation}
(1+h^{\prime }(r)^{2})^{-1/2}h^{\prime }(r)\sim \log r+\mathrm{const.}
\end{equation}
Hence a contradiction because the left-hand-side is bounded in absolute
value by 1 whereas the right-hand-side diverges as $r\searrow 0$. The
contradiction can be seen more concretely as follows: 
\[
h^{\prime }(0)=h^{\prime }(1)-\int_{0}^{1}drh^{\prime \prime }(r)\sim 
\]
\begin{equation}
h^{\prime }(1)-\int_{0}^{1}dr(1+h{^{\prime }}^{2})^{3/2}/r=-\infty ,
\end{equation}
where (\ref{eqq}) was used. This is incompatible with $h\ge 0$ and $h(0)=0$.

We conclude that $\mathbf{t}$ and $\mathbf{n}$ should be continuous all
along the contact line, which therefore should be tangent to the $y$-axis at 
$\pm L/2,\,\pm y_{1},\,\pm y_{2}$ on Fig.~\ref{figL2}. And the contact angle
itself should be continuous, implying $\theta (\pm L/2)=\theta (\pm
y_{1})=\theta _{1}$ and $\theta (\pm y_{2})=\theta _{2}$, forbidding $%
y_{1}=y_{2}$, as in Fig.~\ref{figL2}(c).

Examples of quasi-corners are shown on Fig.~\ref{corner}. They cannot be
droplet equilibrium shapes. The surface shown on Fig.~ \ref{corner}(a) has
mean curvature diverging as $1/r$ as $r\searrow 0$. The surface shown on
Fig.~\ref{corner}(b) has principal curvatures diverging as $\pm 1/(r|\log
r|^{3})$ as $r\searrow 0$, and mean curvature diverging as $|\log r|$ as $%
r\searrow 0$. It is not continuous at the origin, and the geodesic used in
the argument above has a length $\mathcal{O}(1)$ instead of $\mathcal{O}%
(\epsilon )$ as $\epsilon \searrow 0$.

\section{Molecular dynamics simulation}

To study the mechanism behind pinning, we have performed molecular dynamic
(MD) simulations of a liquid drop on top of an inclined solid plate ($\alpha
=30^{\circ }$ with respect the horizontal) in the proximity of a junction
line perpendicular to the inclination. The plate is divided in two
half-planes with different wetting properties (more hydrophilic on top of
the junction and less hydrophilic below) and subject to a vertical force $%
\mathbf{F_{0}}$. See Fig.~\ref{fig:scheme}. Initially, this external force
is equal to zero ($\mathbf{F_{0}}=0$), with the droplet deposited on the
more hydrophilic solid and close to the junction. Once the system reaches
the equilibrium, defined by a constant energy and a constant value of the
local contact angle around the contact line, equal to the equilibrium
contact angle, we introduce a force $\mathbf{{F_{0}}\ne 0}$ acting over all
the liquid atoms. As a consequence, the liquid drop approaches the junction
and the shape of the contact line is altered. Depending on the value of $%
\mathbf{F_{0}}$, three different scenarii are possible in the simulation: $%
\left( a\right) $ the base of the drop can be totally on the hydrophilic
side of the solid, $\left( b\right) $ part of the contact line can cross
over to the hydrophobic solid or $\left( c\right) $ the drop can cross
completely the junction and roll over the hydrophobic substrate. We have
selected a range of $\mathbf{F_{0}}$ to analyze the three regimes. For each
value of $\mathbf{F_{0}}$ used in the simulation, we compute the length of
the intersection between the contact line and the junction ($L$) as well as
the value of the local contact angles at several points $p_{i}$ along the
contact line, $\theta (p_{i})$. Once we have the length $L$ and the contact
angles $\theta (p_{i})$, it is possible to compare the simulation results
and the new versions of the equilibrium equations obtained through the
balance of forces.

\subsection{Setup}

We consider an incline of angle $\alpha =30^{\circ }$ with respect to the
horizontal axis as shown in Fig.~\ref{fig:scheme}. The $x$-axis is along the
slope downhill, the $y$-axis is horizontal. The upper half-plane ($x<0$) is
a relatively hydrophilic substrate $S_{1}$, while the lower half- plane ($%
x>0 $) is a less hydrophilic solid $S_{2}$. The drop profile is $z=h(x,y)$
measured perpendicular to the slope. Close to the $S_{1}/S_{2}$ junction, we
put a liquid droplet. An external vertical force $\mathbf{f}_{0}=f_{0}\sin
\alpha \mathbf{e}_{x}-f_{0}\cos \alpha \mathbf{e}_{z}$ is acting over each
liquid atom and then, the total force acting over the liquid drop is equal
to $\mathbf{F}_{0}=N\mathbf{f}_{0}$ where $N$ is the number of liquid atoms
that compose the droplet. We denote by $F_{0}$ the modulus of $\mathbf{F}%
_{0} $ just like $f_{0}$ is the modulus of $\mathbf{f}_{0}.$

Full details of the simulation methods, base parameters and potentials have
been given in some previous publications (e.g., Ref. \cite
{Coninck08,Bertrand09} and work cited therein). We recall here the key
aspects. The liquids, the solids and their interaction are modeled using
Lennard-Jones potentials defined by: 
\begin{equation}
V(r_{ij})=4\epsilon C_{AB}\left( \left( \frac{\sigma }{r_{ij}}\right)
^{12}-\left( \frac{\sigma }{r_{ij}}\right) ^{6}\right)
\end{equation}

Here, $r_{ij}$ is the distance between any pair of atoms $i$ and $j$. The
coupling parameter $C_{AB}$ enables us to control the relative affinities
between the different types of atoms. The parameters $\epsilon $ and $\sigma 
$ are related, respectively, to the depth of the potential wells and an
effective atomic diameter. For both solid and liquid atoms the diameter $%
\sigma $ is equal to $0.35$ nm, and $\epsilon =k_{B}T$ where $k_{B}$ is the
Boltzmann constant and $T=33$ K is the temperature, which is kept constant
by a thermostat based on velocity scaling. The pair potential is set to zero
for $r_{ij}>2.5\sigma $. $C_{AB}$ is given the value $1$ for both
liquid-liquid (LL) and solid-solid (SS) interactions.

$T=33$ K is indeed a very low temperature, because our simulated liquid is a
simple toy model liquid. In order to model wetting in molecular dynamics
using just Lennard-Jones interactions, we need to work with a system of
atoms in a liquid state with a very low viscosity, able to diffuse in short
periods of time (of the order of ns). {More realistic systems can be
considered but the }time cost of these simulations will be huge. The chosen
parameters for the Lennard-Jones interaction between the linear chains
correspond to a liquid system with a low viscosity and low surface tension
that allows us to study the wetting dynamic in the scale of few nanoseconds
which is reasonable from a computation point of view. This allows us to
understand the mechanism behind the physical process and then, use this
knowledge in a real system where we expect that the same mechanism will be
present. For this particular set of parameters, we have studied drop
spreading, \cite{Coninck08,Bertrand09}, capillary bridges, \cite{BT0}, \cite
{Toledano2017}, wetting of nanofibers, \cite{Seveno2013b}, and in all cases
the behaviour of the simulated liquids mimic well what we can measure in the
laboratory. Then, we consider that a common mechanism is shared between both
systems, the real one and the molecular dynamics simulation. Of course,
realistic simulations can be done to model a specific liquid, but more
complex interactions must be added with effective parameters measured
experimentally. From a fundamental point of view, it is thus easier to work
with the simpler system that has the phenomena that one wants to study.

The solid plate is constructed as $66102$ atoms distributed in a
rectangular, square-planar lattice having three atomic layers whose normal
is parallel to the $z$ axis. The distance between nearest-neighbor solid
atoms is set to $2^{1/6}\sigma $ ($0.393$ nm), i.e., the equilibrium
distance given by the Lennard-Jones potential. The atoms can vibrate around
their initial equilibrium positions according to a harmonic potential
defined by $V_{H}(r)=k|\mathbf{r}-\mathbf{r_{0}}|^{2}$ where $k=1000\epsilon
/\sigma ^{2}$, $\mathbf{r}$ is the instantaneous position of a given solid
atom and $\mathbf{r_{0}}$ its initial position. The plane $x=0$ splits the
solid plate into two half-planes that we use to model the two solid phases
characterized by two different solid-liquid couplings: $C_{LS_{1}}=0.8$ for
the hydrophilic substrate $S_{1}$ which corresponds to a Young angle $\theta
_{1}=72^{\circ }$, while $C_{LS_{2}}=0.4$ is set for the less hydrophilic
solid corresponding to a Young angle $\theta _{2}=135^{\circ }>\theta _{1}$. 
{Then, the value of the local contact angle along the contact line of a drop
trapped between the $S_{1}/S_{2}$ junction will be contained in the interval 
$[72^{\circ },135^{\circ }]$ which allows a large window of analysis.}

The liquid is modeled as $5000$ of $8$-atom molecular chains ($N=40000$
atoms), with adjacent atoms linked by a confining potential $U_{\text{conf}%
}(r_{ij})=Ar_{ij}^{6}$ where $A$ is set to $\epsilon /\sigma ^{6}$. This {%
chain length reduces considerably the evaporation (the vapor phase is
effectively here vacuum) and allows to use more efficiently all the
considered molecules to study pinning }over the time scale of the
simulation. The masses of all the atoms are equated to that of carbon ($12$
g/mol) to allow comparison with physical systems.

The dimensions of the simulation box are $L_{x}=L_{y}=L_{z}=70$ nm and we
impose periodical boundary conditions in the $x$ and $y$ directions. A
sketch of the simulation geometry is shown in Fig.~\ref{fig:scheme}.

Table \ref{tab:1} shows the values for the external force per atom, $f_{0}$,
the total force $F_{0}$ and the corresponding Bond number associated to this
force, $Bo=F_{0}R^{2}/(V\gamma )$, where $V$ is the volume of the liquid
drop and $R$ is the radius of the initial circular contact line, \cite{Rie}.
The range of the Bond numbers showed in this table are similar to the ones
measured by other authors in experiments on the shape and motion of
millimetre-size drops of silicon oil sliding down over an homogeneous plane 
\cite{Grand2005} and of glycerol-water mixtures over a substrate decorated
with linear chemical steps \cite{Semprebon}.

\begin{table}[tbp]
\centering
\begin{tabular}{c|c|c}
$f_0$ ($\times 10^{-3}$ pN) & $F_0=Nf_0$ (pN) & $Bo$ \\ \hline
$0.17$ & $6.64$ & 0.12 \\ 
$0.50$ & $19.92$ & 0.36 \\ 
$0.83$ & $33.21$ & 0.61 \\ 
$11.62$ & $46.48$ & 0.85 \\ 
$16.61$ & $66.41$ & 1.22 \\ 
$19.93$ & $79.69$ & 1.46 \\ 
$24.91$ & $99.62$ & 1.82
\end{tabular}
\caption{Conversion between simulation units and real units for the external
vertical force, and corresponding Bond number.}
\label{tab:1}
\end{table}

In a first step, the molecules of the liquid are distributed in a spherical
region on top of the solid $S_{1}$ and far from the $S_{1}/S_{2}$ junction
meanwhile the temperature of the liquid and the solid is kept constant by
rescaling the atoms velocities. After $10^{6}$ time steps, the system
reaches the equilibrium characterized by a stable value of the energy and a
constant contact angle $\theta _{1}=(70\pm 3)^{\circ }$ along the contact
line defined by the intersection between the liquid, the solid $S_{1}$ and
the vacuum. Then, we introduce an external vertical force $\mathbf{f}_{0}$
acting over all liquid atoms to model a liquid drop. Depending on the
magnitude of this force, the drop can be stuck completely on top of the $%
S_{1}$ solid or it can get pinned in the $S_{1}/S_{2}$ interface where there
appears a segment of length $L$ as the intersection between the contact line
and the junction. As a third possibility, the liquid drop can cross
completely the junction and roll-off over the $S_{2}$ solid. These three
possible behaviors have been observed in our simulations when the value for
the total external force was varied from $6$ (drop stuck inside $S_{1}$) to $%
100$ pN (the limit where the drop rolls off over $S_{2}$). We therefore
select a range of $F_{0}$ between these two limits, $F_{0}=6.6$, $19.9$, $%
33.2$, $46.5$, $66.4$, $79.7$ and $99.6$ pN. For each value of this force we
restart the simulation from the previous equilibrated configuration and we
let the system evolve during $5\times 10^{6}$ time steps, time long enough
to reach a stationary regime characterized by the fluctuation of the energy
and the local contact angles around constant values. After that, we run an
additional $5\times 10^{6}$ time steps where we decompose the liquid in
cubic cells with side $0.1$ nm to calculate the average density inside each
cell every $10^{6}$ time steps to have $5$ independent density computations
for each value of $F_{0}$ that we will use to compute the average values and
their errors for the different magnitudes calculated in this work.

During the application of the external force, the thermostat for scaling of
the velocities is only applied to the solid atoms but the collisions between
the solid and the liquid atoms are enough to maintain constant the liquid
temperature.

\subsection{Contact line and intersection length $L$}

\begin{figure}[h]
\includegraphics[scale=0.19]{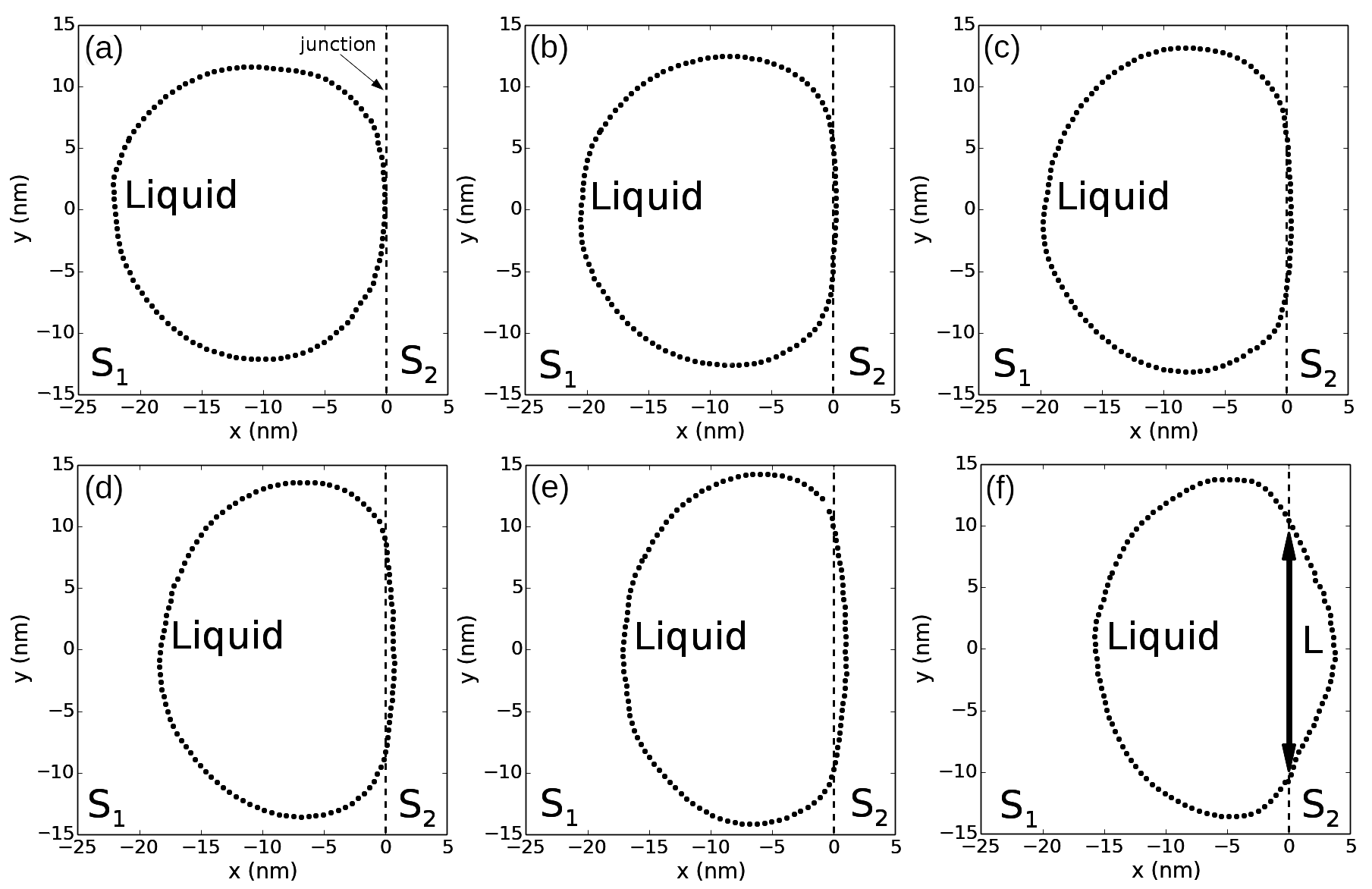} \centering
\caption{Averaged contact line for (a) $F_{0}=6.6$ pN, (b) $F_{0}=33.2$ pN,
(c) $F_{0}=46.5$ pN, (d) $F_{0}=66.4$ pN, (e) $F_{0}=79.7$ pN and (f) $%
F_{0}=99.6$ pN. The dashed line located at $x=0$ represents the location of
the $S_{1}/S_{2}$ junction.}
\label{fig:cline}
\end{figure}

To extract the position of the contact line, it is necessary firstly to
determine the location of the liquid-vacuum ($L/V$) interface that we define
as the locus where the liquid density is $50\%$ of the density of the liquid
in the bulk (i.e. the equimolar surface). To measure the local distribution
of this density, we subdivide the available volume of the drop into cubic
cells of size $dx$, $dy$, $dz=0.3$ nm and calculate the average number of
atoms per cell over $500$ configurations at intervals of $10^{3}$ time steps
to extract $10$ independent density profiles, which we use to establish the
location of the interface, the position of the contact-line, the local
contact angles, and the associated errors. Then, the drop is sliced into $k$
layers parallel to the $L/S$ interface. The density in each slice depends on
the $x$-$y$ coordinates, so we decompose the slice into bins perpendicular
to the $x$ axis and calculate the density profile along the $y$ coordinate.
Finally, this profile is split into two symmetrical regions about the $x$
axis and we fit sigmoidal functions to determine the position of the
interface where the density falls to half that of the bulk. This is done for
each slice in the $z$ direction to locate the complete liquid-vacuum
interface. The contact line is obtained from the intersection between the $%
L/V$ interface and the solid plate. Fig.~\ref{fig:cline} shows the different
equilibrium contact lines for each value of $F_{0}$ used in this work.

Fig.~\ref{fig:cline}(a)(b)(c) clearly corresponds to Fig. {\ref{figL2}}(a).
Fig.~\ref{fig:cline}(f) clearly corresponds to Fig. {\ref{figL2}}(c). Fig.~%
\ref{fig:cline}(d)(e) may correspond to Fig. {\ref{figL2}}(b). The question
is the depth of the protuberance into the hydrophobic side (of order $2$ nm
in the MD simulation) and the tentative $y_{1}y_{2}$ segment (of length also
a few nanometers): do these lengths scale with the size of the drop or
remain of the order of the fluctuations independently of the drop size?

\begin{figure}[tbp]
\begin{center}
\resizebox{8cm}{!}{\includegraphics{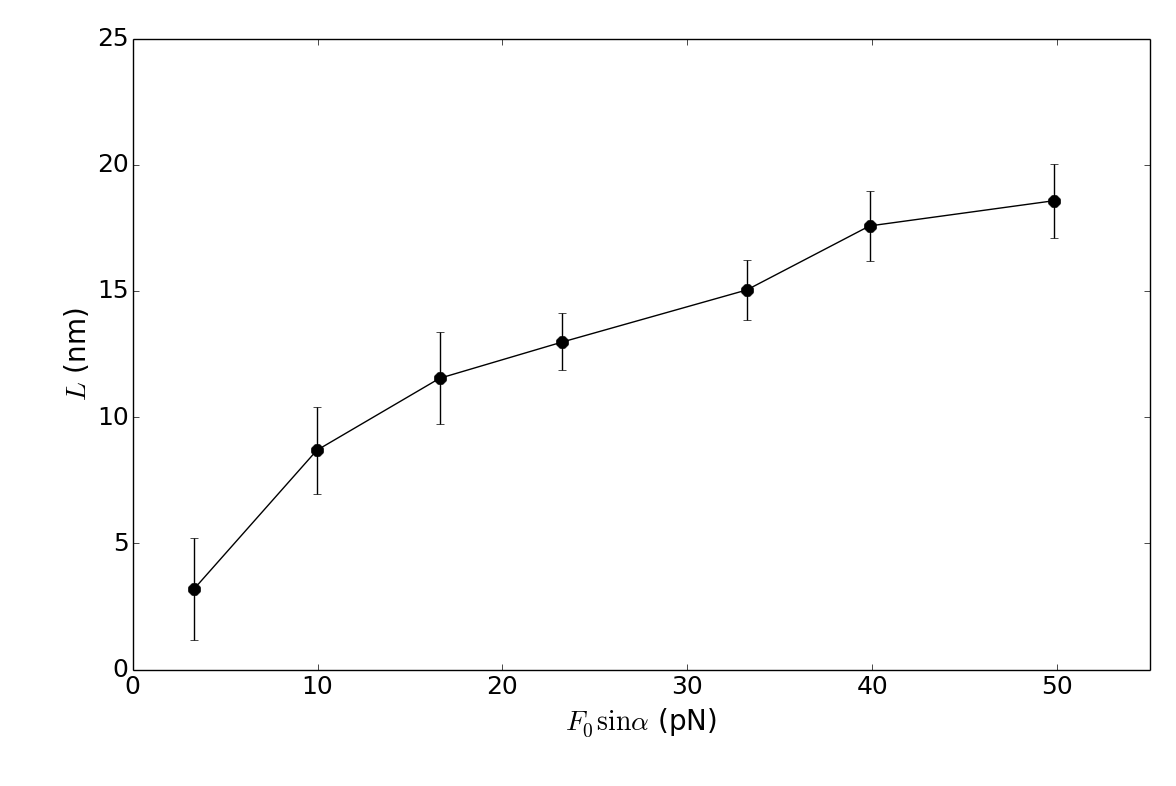}}
\end{center}
\caption{Length $L$ of the segment defined as the intersection between the
contact line and the junction, versus $F_{0}\sin \alpha $.}
\label{fig:L}
\end{figure}

Once the contact line is determined, we compute the intersection between the
contact line and the $S_{1}/S_{2}$ junction located at $x=0$. For this, we
first locate the two pairs of consecutive points of the contact line that
lie on opposite sides with respect to the junction. Then, we determine the
two intersections between the two straight lines defined by each one of
these pairs of points and the junction. From the distance between these two
intersection points we obtain the length of the segment $L$\ shown in Fig.~%
\ref{fig:L}, the values of which are presented in Table \ref{tab:2}.

\subsection{Contact angles}

\begin{figure}[tbp]
\begin{center}
\resizebox{12cm}{!}{\includegraphics{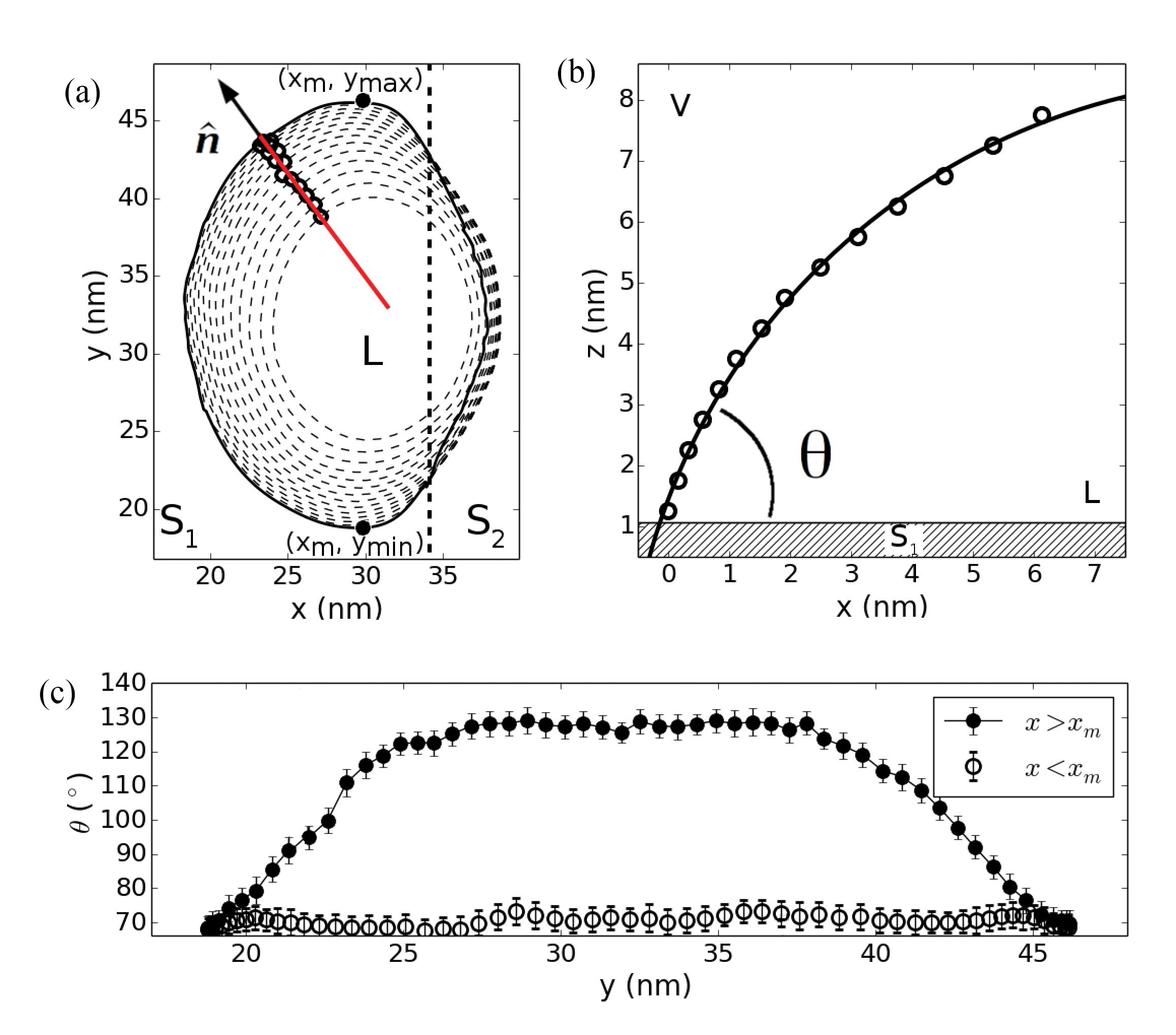}}
\end{center}
\caption{(a) Calculation of the L/V profile at a given point of the contact
line. The extremes of the contact line along the $y$-axis are located at $%
x=x_{m}$. (b) L/V profile and circular fitting used to obtain the local
contact angle. (c) Local contact angle versus $y$ for $x<x_{m}$ (open
symbols) and $z>x_{m}$ (full symbols). All figures correspond to $F_{0}=99.6$
pN.}
\label{fig:calcCA}
\end{figure}

\begin{figure}[tbp]
\begin{center}
\resizebox{10cm}{!}{\includegraphics{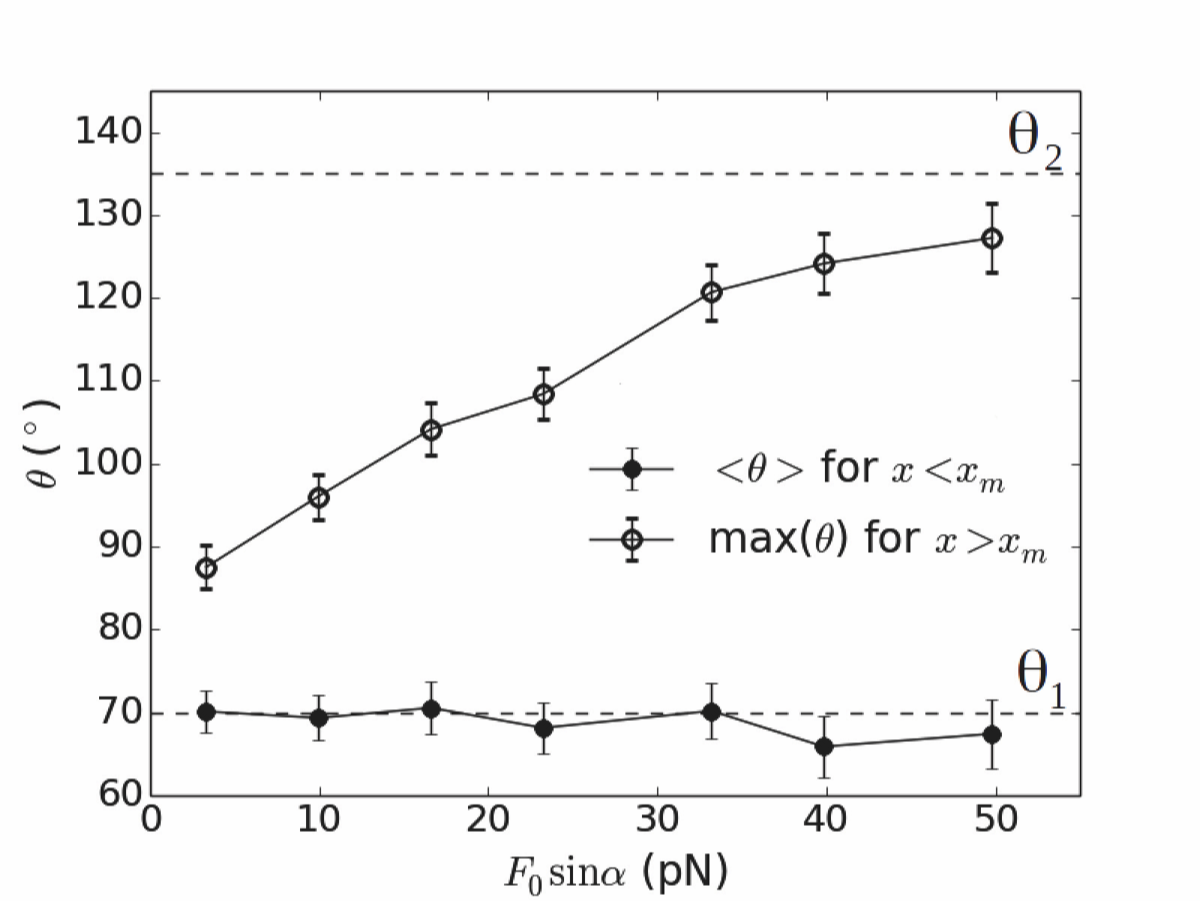}}
\end{center}
\caption{Average value of the local contact angle for $x<x_{m}$, and maximum
of local contact angle along the contact line for $x>x_{m}$, versus $%
F_{0}\sin \alpha $.}
\label{fig:th1th2}
\end{figure}

For the computation of the local contact angles, we calculate the normal to
the contact line in each one of the points of this contact line. Then, we
compute the intersection of translates of this normal line with the $L/V$
interface at different heights $z$ as it is sketched in Fig.~\ref{fig:calcCA}%
(a). This gives us the profile associated to each point $p_{i}$ of the
contact line. We fit a circular arc to this profile and from the slope of
this fitted circle at its intersection with the solid we compute the local
contact angle at the point $p_{i}$, $\theta (p_{i})$, as it is shown in Fig.~%
\ref{fig:calcCA}(b). The extremes in $y$ along the contact line are located
at $x=x_{m}$, $y=\pm y_{m}$; we use this to split the contact line in two
sets of points for which $x>x_{m}$ and $x<x_{m}$.

Fig.~\ref{fig:calcCA}(c) shows an example of the dependence of the local
contact angle with the $y$ coordinate for $x<x_{m}$ and $x>x_{m}$ for $%
F_{0}=99.6$ pN. The contact line points at $x<x_{m}$ are located on top of
the solid $S_{1}$ and they exhibit a constant contact angle equal to $\theta
_{1}$, i.e., to the equilibrium contact angle between this liquid and $S_{1}$%
. However, for $x>x_{m}$ the contact angle depends on $y$ and varies between 
$\theta _{1}$ and a maximum value $\theta _{\max }$. As the external force
is increased, $\theta _{\max }$\ also increases until it reaches the maximum
value $\theta _{\max }=\theta _{2}$,\ as can be seen in Table \ref{tab:2}.
The average value of the contact angle for $x<x_{m}$ and its maximum value
for $x>x_{m}$ are shown in Fig.~\ref{fig:th1th2}.

In cases (b)(c), $\theta _{\max }=\theta _{2}$ while in case (a), $\theta
_{\max }$ depends upon the Bond number and is significantly less than $%
\theta _{2}$ in agreement with the data in Fig. \ref{fig:th1th2} and Fig. 
\ref{fig:integral} for $F_{0}\sin \alpha <30$ pN.

As can easily be observed in Fig. \ref{fig:th1th2}, when the applied force
is small enough, we do observe a significant difference between $\theta _{2}$%
\ and $\theta _{\max }$. This is due to the fact that the corresponding
contact line is in the vicinity of the $S_{1}/S_{2}$\ junction, leading to a
clear modification of the density of the liquid in contact with the solid.
Indeed, we do observe an increase of the density $\rho _{S_{2}}$ of the
liquid able to cross the $S_{1}/S_{2}$\ junction, as $F_{0}$\ is increased
until it reaches the characteristic density of the liquid deposited on top
of an homogeneous $S_{2}$\ solid ($\rho _{S_{2}}^{0}=21.4$\ atoms/nm$^{3}$).
This can be seen in Table \ref{tab:2} where $\rho _{S_{2}}<\rho _{S_{2}}^{0}$%
\ and $\theta _{\max }<\theta _{2}$.

\begin{table}[tbp]
\centering
\begin{tabular}{c|c|c|c}
$F_0\sin\alpha$ (pN) & $\theta_{\max}$ ($^\circ$) & $L$ (nm) & $\rho_{S_2}$%
(atoms/nm$^3 $) \\ \hline
$3.32$ & 87.5 & 3.6 & 0.6 \\ 
$9.96$ & 96.1 & 9.7 & 3.4 \\ 
$16.61$ & 104.1 & 12.8 & 6.9 \\ 
$23.24$ & 108.4 & 14.4 & 11.1 \\ 
$33.21$ & 120.7 & 16.7 & 18.0 \\ 
$39.85$ & 124.1 & 19.5 & 19.6 \\ 
$49.81$ & 127.2 & 20.2 & 20.1
\end{tabular}
\caption{$\theta_{\max}$, $L$ and density of the first layer of liquid on
top of the solid $S_2$ for the different values of the external force $F_0$
considered in this work. The value of this density if the liquid is totally
immersed in $S_2$ (no $S_1$/$S_2$ junction) is $\rho_{S_2}^0=21.4$ atoms/nm$%
^3 $.}
\label{tab:2}
\end{table}

\section{Comparison between simulation results and model}

\begin{figure}[tbp]
\centering
\includegraphics[scale=0.4]{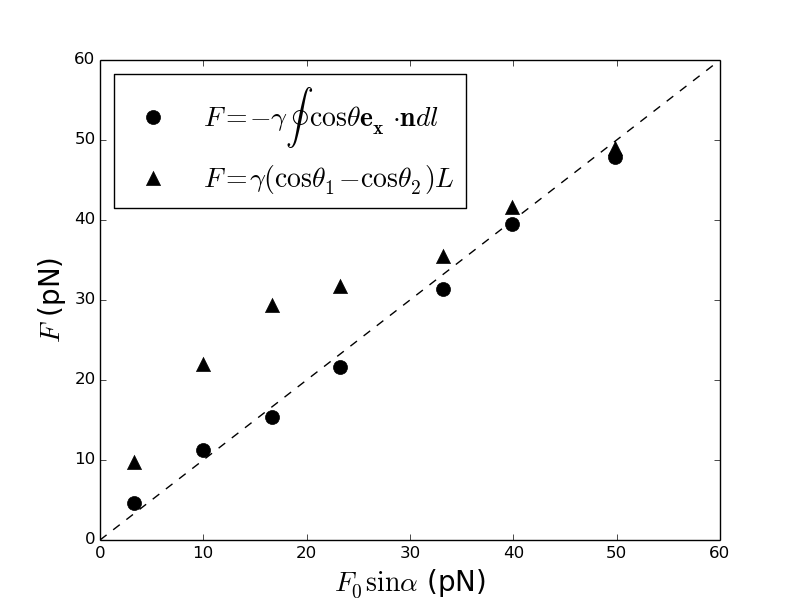}
\caption{Modulus $F$ of downhill component of capillary force upon drop,
versus downhill component $F_{0}\sin \alpha $ of volume (weight) force.
Circles: numerical integration of Eq. (\ref{capf}) projected onto the $x$%
-axis, corresponding to Eq. (\ref{eqx}). Triangles: heuristic formula in
terms of wetted length $L$ of junction line, corresponding to Eq. (\ref{eqc}%
). The dashed line represents the function $F=F_{0}\sin \alpha $.}
\label{fig:integral}
\end{figure}

\begin{figure}[tbp]
\centering
\includegraphics[scale=0.4]{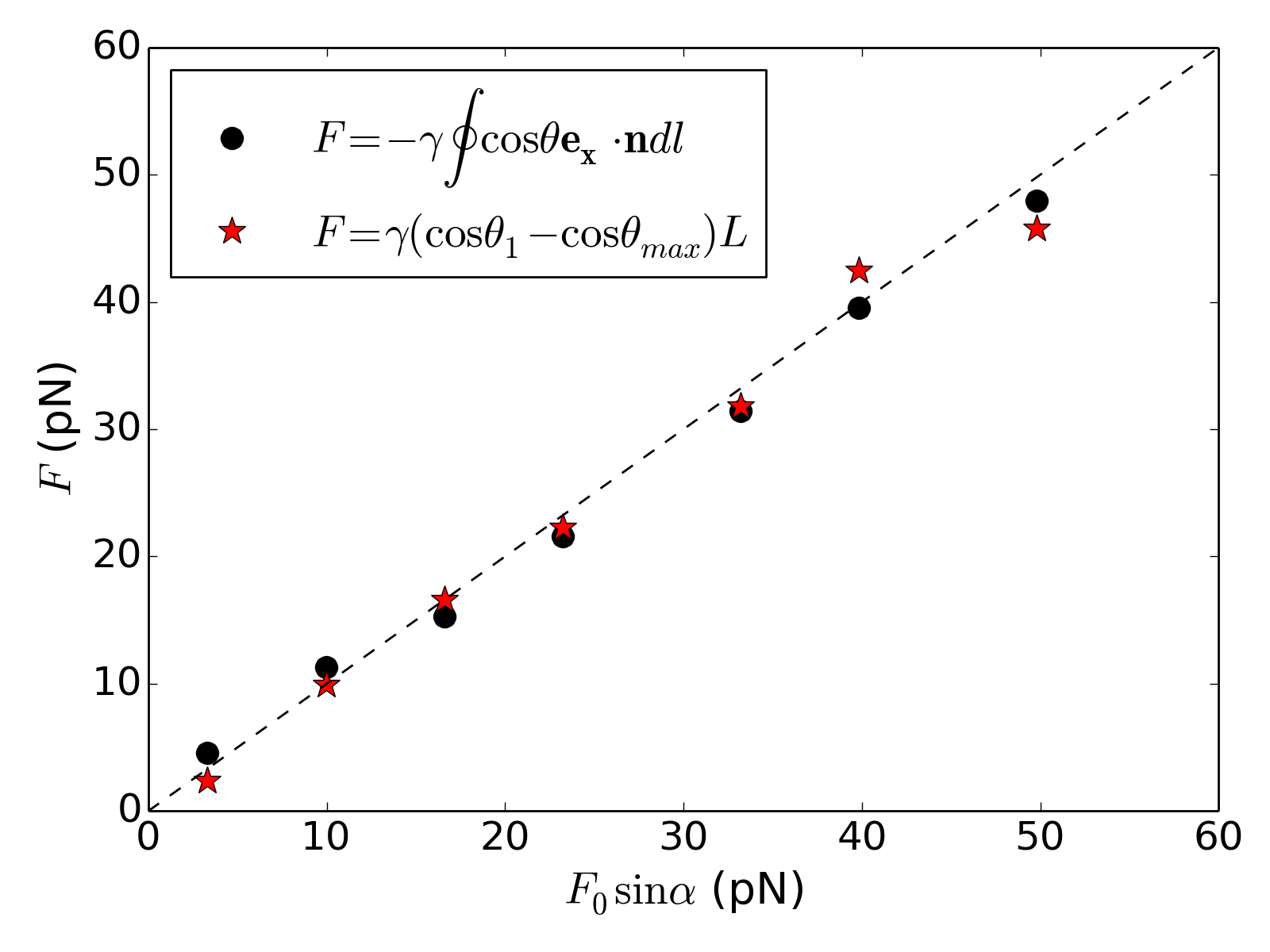}
\caption{Modulus $F$ of downhill component of capillary force upon drop,
versus downhill component $F_{0}\sin \alpha $ of volume (weight) force.
Circles: same as in Fig.~\ref{fig:integral}. Stars: heuristic formula in
terms of wetted length $L$ of junction line, using the maximum $\theta
_{max} $ of the contact angle along the contact line, approximating Eq. (\ref
{eqa}). }
\label{fig:integral2}
\end{figure}

Once we have measured the distribution of the local contact angle along the
contact line, it is possible to check the force balance parallel to the
Solid-Liquid interface, as given by Eq. (\ref{eqx}). To do so, we compute
numerically the integral appearing in the modulus $F=-\gamma
\,\oint_{\partial \Omega }\,dl\,\mathbf{n}\cdot \mathbf{e_{x}}\,\cos \theta $
of the downhill component of capillary force upon the drop, using a simple
trapezoidal method. In practice, we select a set of points around the
contact line where we compute the value of the local contact angle $\theta $
and the term $\mathbf{n}\cdot \mathbf{e_{x}}$ inside the contour integral.
Fig.~\ref{fig:integral} shows $F$ versus $F_{0}\sin \alpha $, the downhill
component of volume (weight) force. We observe a very good agreement between
the simulation results and the model in the full range. Therefore, it is
clear that the general Eq. (\ref{eqx}) can be used to determine the total
force parallel to the Solid-Liquid interface knowing a distribution of
points along the contact line and the local contact angle on these points
for any value of the external force.

The limitation of the system size in our MD simulations will inevitably
affect the amplitude of the contact line fluctuations. It will therefore be
difficult to identify case Fig.~\ref{figL2}(b) and distinguish it from the
end of case Fig.~\ref{figL2}(a) or the beginning of case Fig.~\ref{figL2}%
(c). When the value of $F_{0}$ is high enough to reach the unstable
configuration sketched in Fig.~\ref{figL2}(c), it is possible to use the
simplified equation $L\gamma (\cos \theta _{1}-\cos \theta _{2})$ (Eq. (\ref
{eqc})) where the only needed parameters are the liquid-vapor surface
tension ($\gamma $), the length of the intersection of the contact line with
the junction ($L$) and the values of the Young angles of the liquid
deposited on top of each one of the solids ($\theta _{1}$ and $\theta _{2}$%
). Fig.~\ref{fig:integral} shows that this simple model agrees remarkably
well for values of $F_{0}\sin \alpha \gtrsim 33$ pN which corresponds to the
system represented in Fig.~\ref{fig:cline}(d), (e) and (f) with $%
F_{0}\gtrsim 66$ pN. Here the advancing contact angle approaches the Young
angle of the liquid on the more hydrophobic solid $S_{2}$ as it can be seen
in Fig.~\ref{fig:th1th2} and then, we are in the situation sketched in Fig.~%
\ref{figL2}(c) where Eq. (\ref{eqc}) should hold. Finally, in Fig.~\ref
{fig:integral2}, it is shown that in the range $F_{0}\sin \alpha \lesssim 33$
pN, corresponding to the situations sketched in Fig.~\ref{fig:cline}(a), (b)
and (c) with $F_{0}\lesssim 66$ pN, it is possible to use the simplified
equation $L\gamma (\cos \theta _{1}-\cos \theta _{\max })$ involving the
maximum value $\theta _{\max }$ of the contact angle of the drop along the
contact line. Concerning Fig. \ref{fig:integral2} in the thermodynamic
limit, the formula for the black circles is exact while the formula for the
red stars comes from the approximation $\theta =\theta _{\max }$ all along
the part of the contact line with $x\geq 0$. Concerning Fig. \ref
{fig:integral}, the formula for the black triangles is exact in case (c)
where $F_{0}\sin \alpha >80$ (not shown on the figure), assuming local
equilibrium at any point on the contact line.

\section{Conclusion}

We have studied a drop pinned on an incline at the junction between a
hydrophilic half-plane and a hydrophobic one. In spite of the discontinuity
of the $S/L$ surface tension at the junction, we have shown that the contact
line must remain a differentiable curve. Based on the equilibrium equations
derived from the balance of forces, we have described theoretically three
different scenarii: (a) one for which the contact line partly follows the
junction line on a segment of width $L$, (b) one for which part of the
contact line goes into the hydrophobic half-plane in a central protuberance
while keeping two side overlaps with the junction line and (c) one for which
part of the contact line crosses straight into the hydrophobic half-plane
(See Fig. {\ref{figL2}}).

In all three cases, we find a formula in the spirit of Furmidge formula Eq. (%
\ref{eq:Fermigier}), namely:

\begin{equation}
mg\sin \alpha \simeq \gamma \,L\,(\cos \theta _{1}-\cos \theta _{\max }),
\label{Furmg}
\end{equation}
where $L$ is the wetted length of the junction line and $\theta _{\max }$ is
the maximum of the contact angle along the contact line, which is also the
contact angle at the front of the drop. In case (a): $\theta _{\max }<\theta
_{2}$ and $\theta _{\max }$ depends upon the Bond number. In cases (b)(c): $%
\theta _{\max }=\theta _{2}$.

To check the validity of the exact formula (\ref{eqx}) and the approximate
formula (\ref{Furmg}), we have performed molecular dynamics simulations for
different values of the gravity force $F_{0}=mg.$ We then extracted the
local contact angle along the contact line as well as the length $L$ of the
contact line intersection with the junction. Then, we have used them to
verify the exact full force balance given by Eq. (\ref{eqx}) showing
excellent agreement. We also checked MD simulation results against the
approximate formula (\ref{Furmg}) in a wide range of $F_{0}=mg$ showing
again an excellent agreement. We find a range of small $F_{0}$ where
scenario (a) is observed and $\theta _{\max }<\theta _{2}$, then a medium
range of $F_{0}$ where scenario (b) is observed and $\theta _{\max }$
slightly smaller than $\theta _{2}$ and a range of larger $F_{0}$ where
scenario (c) is observed and $\theta _{\max }=\theta _{2}$.\newline

\textbf{Acknowledgement.} This research was partially funded by the
Inter-University Attraction Poles Programme (IAP 7/38 MicroMAST) of the
Belgian Science Policy Office. The authors also thank FNRS and RW for
partial support. Computational resources have been provided by the
Consortium des Equipements de Calcul Intensif (CECI), funded by the Fonds de
la Recherche Scientifique de Belgique (F.R.S.-FNRS) under Grant No.
2.5020.11.

\end{document}